\pdfoutput=1
\documentclass[
  reprint,
  superscriptaddress,
 amsmath,amssymb,
 aps,
 prl
]{revtex4-1}

\usepackage{graphicx}
\usepackage{dcolumn}
\usepackage{bm}
\usepackage{color}

\begin{document}

\title{
  Light-induced large and
  tunable valley-selective Hall effect
  in a centrosymmetric system
}

\author{Naoya Arakawa}
\email{arakawa@phys.chuo-u.ac.jp}
\affiliation{The Institute of Science and Engineering,
  Chuo University, Bunkyo, Tokyo, 112-8551, Japan}
\author{Kenji Yonemitsu}
\affiliation{The Institute of Science and Engineering,
  Chuo University, Bunkyo, Tokyo, 112-8551, Japan}
\affiliation{Department of Physics,
  Chuo University, Bunkyo, Tokyo 112-8551, Japan}


\begin{abstract}
  We propose that
  a large and tunable
  valley-selective Hall effect can be realized in a centrosymmetric system
  via light-induced breaking of inversion and time-reversal symmetries.
  This is demonstrated in graphene driven by bicircularly polarized light,
  which consists of a linear combination of left- and right-handed
  circularly polarized light with different frequencies.
  We also show
  that
  our Hall conductivity is two orders of magnitude larger than
  the maximum value obtained in noncentrosymmetric systems, and that 
  the main valley can be switched 
  by tuning a phase difference between the left- and right-handed circularly polarized light. 
  Our results will enable us to generate and control the valley-selective Hall effect
  in centrosymmetric systems.
\end{abstract}
\maketitle


\textit{Introduction.}Electron systems can get a valley degree of freedom
with broken inversion symmetry.
In epitaxial or bilayer graphene~\cite{Epi-graphene,Bilayer-graphene} and
some transition-metal dichalcogenides~\cite{TMD-valley},
electrons at the K and K$^{\prime}$ points in the momentum space
have the valley degree of freedom.
These systems are noncentrosymmetric because
their lattices break inversion symmetry;
this symmetry holds in centrosymmetric systems. 
The valley degree of freedom can be used to realize 
valleytronics phenomena. 
An example is the valley-selective Hall effect,
in which the current from one of two valleys causes
the charge current perpendicular to an electric field.
This was experimentally observed
in MoS$_{2}$~\cite{VHE-exp1} or bilayer graphene~\cite{VHE-exp2}
with resonant circularly polarized light (CPL),
which excites an electron at one valley~\cite{Valley-CPL-Graphene,Valley-CPL-TMD1,Valley-CPL-TMD2,Valley-CPL-TMD3}.
Another example is a valley-contrasting Hall effect~\cite{VHE-theory},
in which the electrons around different valleys generate opposite currents
perpendicular to the electric field.
This is distinct from the valley-selective Hall effect
because in the latter the contribution from one valley is negligible. 
Since the valley degree of freedom can be utilized in similar ways to
the spin degree of freedom in spintronics,
valleytronics has opened 
new phenomena and applications utilizing it. 

\begin{figure}
  \includegraphics[width=80mm]{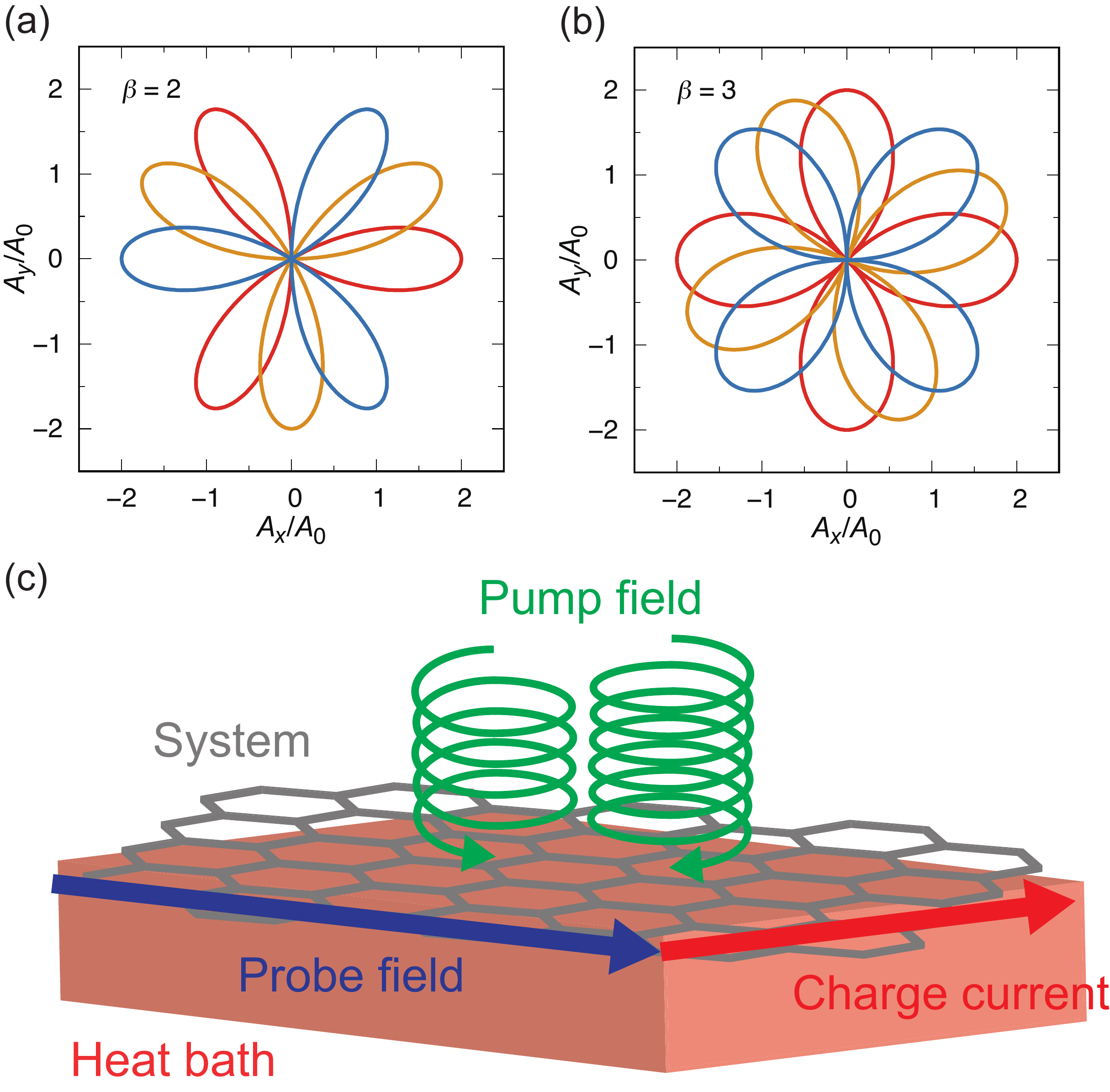}
  \caption{\label{fig1}
    (a), (b) Trajectories of $\mathbf{A}_{\textrm{BCPL}}(t)$
    per period $T_{\textrm{p}}=2\pi/\Omega$ 
    for $\beta=2$ and $3$.
    The red, yellow, and blue lines correspond to those at $\theta=0$, $\frac{\pi}{2}$, and $\pi$,
    respectively.
    (c) The set-up for the valley-selective Hall effect
    in graphene driven by BCPL.
    The system, graphene, is driven by the pump field, the field of BCPL,
    and is weakly coupled to the heat bath.
    The charge current perpendicular to the probe field is generated.
  }
\end{figure}

Bicircularly polarized light (BCPL) will provide a new way
for generating and controlling the valley degree of freedom.
BCPL is generated by a linear combination of left- and right-handed CPL~\cite{Bicirc-NatPhoto}
and described by $\mathbf{A}_{\textrm{BCPL}}(t)=(A_{x}(t)\ A_{y}(t))^{T}$
[e.g., see Figs. \ref{fig1}(a) and \ref{fig1}(b)], where 
\begin{align}
  A_{x}(t)+iA_{y}(t)
  =A_{0}e^{i\Omega t}+A_{0}e^{-i(\beta\Omega t-\theta)}.\label{eq:BCPL}
\end{align}
If $A_{0}$ is strong enough to be treated nonperturbatively,
BCPL can break time-reversal symmetry and inversion symmetry~\cite{Bicirc-Oka,Bicirc-PRL}.
Therefore, 
BCPL could generate
the valley degree of freedom. 
To treat such nonperturbative effects,
we need to consider a system driven by BCPL using the Floquet theory~\cite{Floquet1,Floquet2}.
Since the Hamiltonian in this theory can be changed 
by varying parameters of $\mathbf{A}_{\textrm{BCPL}}(t)$, 
BCPL could also control the valley degree of freedom.
However,
a possibility of a BCPL-induced valley-selective or valley-contrasting Hall effect 
remains unexplored.
    
\begin{figure*}
  \includegraphics[width=160mm]{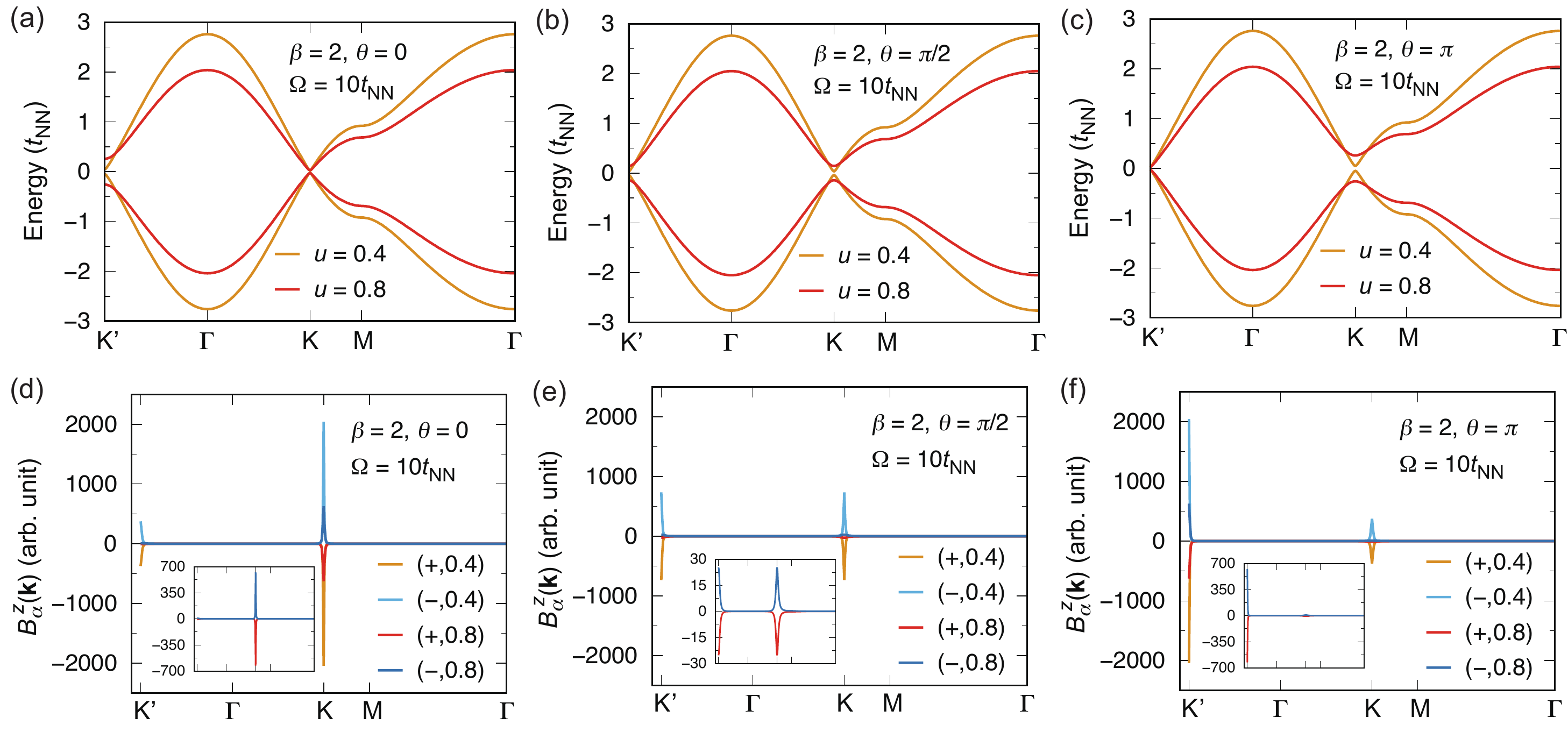}
  \caption{\label{fig2}
    (a){--}(c)
    The momentum dependences of 
    $\epsilon_{+}^{(\textrm{eff})}(\mathbf{k})$ and $\epsilon_{-}^{(\textrm{eff})}(\mathbf{k})$
    obtained in the high-frequency expansion
    for $\beta=2$ at $\Omega=10t_{\textrm{NN}}$, $u=0.4$ and $0.8$,
    and $\theta=0$, $\frac{\pi}{2}$, and $\pi$.
    The momenta at the symmetric points K$^{\prime}$, $\Gamma$, K, and M
    are $(\frac{8\pi}{3\sqrt{3}}\ 0)^{T}$,
    $(0,0)^{T}$,
    $(\frac{4\pi}{3\sqrt{3}}\ 0)^{T}$,
    and $(\frac{\pi}{\sqrt{3}}\ \frac{\pi}{3})^{T}$,
    respectively.
    Here $(\frac{8\pi}{3\sqrt{3}}\ 0)^{T}$
    and $(0,0)^{T}$ are 
    equivalent to
    $(-\frac{4\pi}{3\sqrt{3}}\ 0)^{T}$ and
    $(\frac{4\pi}{\sqrt{3}}\ 0)^{T}$,
    respectively.
    (d){--}(f) The momentum dependences of
    $B_{\alpha}^{z}(\mathbf{k})$ obtained in the high-frequency expansion
    for $\beta=2$ at $\Omega=10t_{\textrm{NN}}$
    and $\theta=0$, $\frac{\pi}{2}$, and $\pi$
    with $(\alpha,u)=(+,0.4)$, $(-,0.4)$, $(+,0.8)$, and $(-,0.8)$.
    The insets show the results for $u=0.8$ on a smaller scale.
    }
\end{figure*}

Here we demonstrate the large and tunable
valley-selective Hall effect in monolayer graphene driven by BCPL.
Using a high-frequency expansion~\cite{Highw1,Highw2} of the Floquet theory,
we show that 
inversion symmetry can be broken only for even $\beta$ in Eq. (\ref{eq:BCPL}),
whereas time-reversal symmetry can be broken for any $\beta$.
We also show the effects of BCPL on
the energy bands and the Berry curvatures.
Then,
using
the Floquet linear-response theory~\cite{Eckstein,Oka-PRB,Tsuji,Mikami,NA-FloquetSHE,NA-linear},
we show that
the valley-selective Hall effect can be realized for $\beta=2$ at $\theta=0$ or $\pi$.
Our Hall conductivity is much larger than the maximum value known so far,
and 
the main valley can be switched by tuning $\theta$.
Our results indicate that
BCPL can be used to generate and control the valley-selective Hall effect
in centrosymmetric systems.
This work will open the door to the valley-selective Hall effect
in centrosymmetric systems.

\textit{Model.}Graphene driven by BCPL
[Fig. \ref{fig1}(c)] is described by the Hamiltonian, 
\begin{align}
  H=H_{\textrm{s}}(t)+H_{\textrm{b}}+H_{\textrm{sb}},\label{eq:H}
\end{align}
where 
$H_{\textrm{s}}(t)$ is the system Hamiltonian for graphene
with the BCPL field $\mathbf{A}_{\textrm{BCPL}}(t)$,
$H_{\textrm{b}}$ is the bath Hamiltonian,
and $H_{\textrm{sb}}$ is the system-bath Hamiltonian.
We have treated the nonperturbative effects of the BCPL field
as the Peierls phase factors of
the kinetic energy~\cite{Oka-PRB,Mikami,NA-FloquetSHE,NA-linear}.
In addition to $H_{\textrm{s}}(t)$,
we have considered $H_{\textrm{b}}$ and $H_{\textrm{sb}}$~\cite{Tsuji,Mikami,NA-FloquetSHE,NA-linear},
where the bath is B\"{u}ttiker-type~\cite{HeatBath1,HeatBath2}
and in equilibrium at temperature $T$.
The main effect of these Hamiltonians is
to induce the damping $\Gamma$~\cite{Tsuji,Mikami,NA-FloquetSHE},
which could be used to realize a nonequilibrium steady state
under heating~\cite{Heating1,Heating2} induced by BCPL.
Throughout this paper,
we set $\hbar=k_{\textrm{B}}=c=a_{\textrm{NN}}=1$,
where $a_{\textrm{NN}}$ is the length between nearest neighbor sites.

\textit{Inversion or time-reversal symmetry breaking.{--}}We begin with
symmetry breaking induced by BCPL. 
Using the high-frequency expansion~\cite{Highw1,Highw2}
of the Floquet theory, 
we obtain an effective Hamiltonian for describing
the nonperturbative effects of off-resonant BCPL,
\begin{align}
  H_{\textrm{eff}} = \sum_{\mathbf{k}}\sum_{a,b=A,B}\sum_{\sigma=\uparrow,\downarrow}
  \bar{\epsilon}_{ab}(\mathbf{k})
  c_{\mathbf{k} a\sigma}^{\dagger}c_{\mathbf{k} b\sigma},\label{eq:Heff}
\end{align}
where
\begin{align}
  &\bar{\epsilon}_{AA}(\mathbf{k})=-\bar{\epsilon}_{BB}(\mathbf{k})
  =\Delta_{\textrm{BCPL}}+K_{AA}^{(\textrm{eff})}(\mathbf{k}),\label{eq:e_AA^eff}\\
  &\bar{\epsilon}_{AB}(\mathbf{k})=\bar{\epsilon}_{BA}(\mathbf{k})^{\ast}
  =\epsilon_{AB}^{(\textrm{eff})}(\mathbf{k}).
  \label{eq:e_AB^eff}
\end{align}
(For the derivation, see the Supplemental Material~\cite{SM}.)
Here 
$\Delta_{\textrm{BCPL}}$ is the staggered sublattice
potential~\cite{Graphene-ISB-potential,Kane-Mele,Bicirc-Oka},
$\epsilon_{AB}^{(\textrm{eff})}(\mathbf{k})=\sum_{\mathbf{R}=\mathbf{R}_{0},\mathbf{R}_{1},\mathbf{R}_{2}}
t_{\textrm{eff}}^{(AB)}(\mathbf{R})e^{-i\mathbf{k}\cdot\mathbf{R}}$,
and
$K_{AA}^{(\textrm{eff})}(\mathbf{k})
=\sum_{\mathbf{R}=\pm\mathbf{R}^{\prime}_{0}, \pm\mathbf{R}^{\prime}_{1}, \pm\mathbf{R}^{\prime}_{2}}
K_{\textrm{eff}}^{(AA)}(\mathbf{R})e^{-i\mathbf{k}\cdot\mathbf{R}}$,
where $t_{\textrm{eff}}^{(AB)}(\mathbf{R})$ and $K_{\textrm{eff}}^{(AA)}(\mathbf{R})$
are the nearest-neighbor and next-nearest-neighbor hopping integrals, respectively.
Note that 
$\mathbf{R}_{0}=(0\ 1)^{T}$,
$\mathbf{R}_{1}=(-\frac{\sqrt{3}}{2}\ -\frac{1}{2})^{T}$,
$\mathbf{R}_{2}=(\frac{\sqrt{3}}{2}\ -\frac{1}{2})^{T}$,
$\mathbf{R}^{\prime}_{0}=(\sqrt{3}\ 0)^{T}$,
$\mathbf{R}^{\prime}_{1}=(-\frac{\sqrt{3}}{2}\ \frac{3}{2})^{T}$,
and $\mathbf{R}^{\prime}_{2}=(-\frac{\sqrt{3}}{2}\ -\frac{3}{2})^{T}$
(see Fig. 1 of the Supplemental Material~\cite{SM}). 
Equations (\ref{eq:Heff}){--}(\ref{eq:e_AB^eff}) show that 
BCPL not only modifies
the nearest-neighbor hopping integrals,
but also induces the staggered sublattice potential and
the next-nearest-neighbor hopping integrals.
The latter quantities depend on $\beta$ [see Eqs. (34) and (35)
  in the Supplemental Material~\cite{SM}]: 
for even $\beta$,
$\Delta_{\textrm{BCPL}}$ is finite
and $K_{\textrm{eff}}^{(AA)}(\mathbf{R})$ has
real and imaginary parts;
for odd $\beta$,
$\Delta_{\textrm{BCPL}}$ is zero
and $K_{\textrm{eff}}^{(AA)}(\mathbf{R})$ is pure-imaginary.
As specific cases,
we show them for $\beta=2$ and $3$ in $u=eA_{0}\ll 1$
(for their derivation, see the Supplemental Material~\cite{SM}):
\begin{align}
  &\Delta_{\textrm{BCPL}}
  =\begin{cases}
  -6K_{1}(u,\theta)\ (\beta = 2),\\
  0\ \ \ \ \ \ \ \ \ \ \ \ \ \ (\beta = 3),
  \end{cases}\label{eq:Del-beta}\\
  &K_{\textrm{eff}}^{(AA)}(\pm\mathbf{R}_{l}^{\prime})
  =\begin{cases}
  \mp iK_{0}(u)+K_{1}(u,\theta)\ \ \ \ \ (\beta = 2),\\
  \mp i[K_{0}^{\prime}(u)+K_{l+1}^{\prime}(u,\theta)]\ (\beta = 3),
  \end{cases}\label{eq:K-beta}
\end{align}
where
$K_{0}(u)=\frac{\sqrt{3}t_{\textrm{NN}}^{2}}{4\Omega}u^{2}$,
$K_{1}(u,\theta)=\frac{t_{\textrm{NN}}^{2}}{4\Omega}u^{3}\cos\theta$,
$K_{0}^{\prime}(u)=\frac{\sqrt{3}t_{\textrm{NN}}^{2}}{4\Omega}(u^{2}-\frac{3}{4}u^{4})$,
$K_{1}^{\prime}(u,\theta)=\frac{\sqrt{3}t_{\textrm{NN}}^{2}}{16\Omega}u^{4}\cos\theta$, 
$K_{2}^{\prime}(u,\theta)
=\frac{t_{\textrm{NN}}^{2}}{16\Omega}u^{4}[\sin\theta-\sin(\theta-\frac{4\pi}{3})]$,
and $K_{3}^{\prime}(u,\theta)
=-\frac{t_{\textrm{NN}}^{2}}{16\Omega}u^{4}[\sin\theta-\sin(\theta-\frac{2\pi}{3})]$.
Therefore, 
BCPL can break inversion symmetry only for even $\beta$
because it is broken by 
$\Delta_{\textrm{BCPL}}$ and the real parts
of the next-nearest-neighbor hopping integrals.
Furthermore,
since time-reversal symmetry is broken by 
the imaginary parts of
the next-nearest-neighbor hopping integrals~\cite{Haldane,Kitagawa,Mikami},
BCPL can break it for any $\beta$.
Note that
the $l$-independent $K_{\textrm{eff}}^{(AA)}(\pm\mathbf{R}_{l}^{\prime})$'s for $\beta=2$
preserve $C_{3}$ rotational symmetry,
whereas the $l$-dependent ones for $\beta=3$
break it.

\begin{figure}
  \includegraphics[width=86mm]{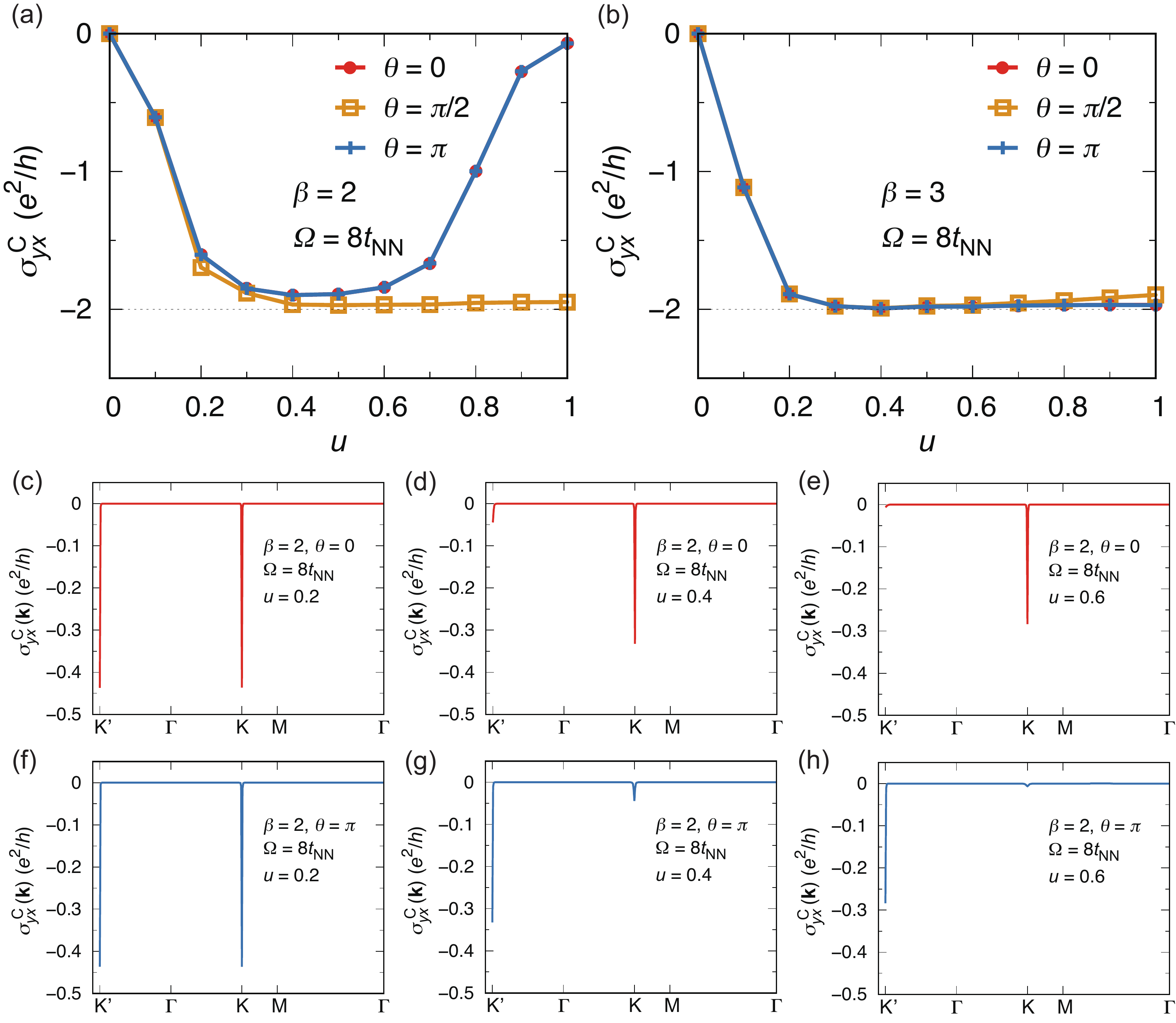}
  \caption{\label{fig3}
    (a), (b) The $u$ dependences of $\sigma_{yx}^{\textrm{C}}$
    obtained in the Floquet linear-response theory
    for $\beta=2$ or $3$ at $\Omega=8t_{\textrm{NN}}$
    and $\theta=0$, $\frac{\pi}{2}$, and $\pi$.
    Here $u=eA_{0}$ is dimensionless. 
    The dotted lines correspond to a quantized value $-2e^{2}/h$.
    (c){--}(h) The momentum dependences of $\sigma_{yx}^{\textrm{C}}(\mathbf{k})$
    obtained in the Floquet linear-response theory
    for $\beta=2$ at $\Omega=8t_{\textrm{NN}}$
    with $(\theta,u)=(0,0.2)$, $(0,0.4)$, $(0,0.6)$,
    $(\pi,0.2)$, $(\pi,0.4)$, and $(\pi,0.6)$.
    The symmetric points are the same as those used in Fig. \ref{fig2}.
    }
\end{figure}

\textit{Valley degeneracy lifting.}The above differences between
the effects of BCPL for even and odd $\beta$ lead to
a difference in the valley degeneracy.
The energy dispersion of Eq. (\ref{eq:Heff}) is given by
\begin{align}
  \epsilon_{\pm}^{(\textrm{eff})}(\mathbf{k})
  =\pm\sqrt{[\Delta_{\textrm{BCPL}}+K_{AA}^{(\textrm{eff})}(\mathbf{k})]^{2}
    +|\epsilon_{AB}^{(\textrm{eff})}(\mathbf{k})|^{2}}.
\end{align}
We estimate $\epsilon_{\pm}^{(\textrm{eff})}(\mathbf{k})$'s
at the K and K$^{\prime}$ points for $\beta=2$ and $3$ in $u\ll 1$;
the results for $\beta=2$ are 
$\epsilon_{\pm}^{(\textrm{eff})}(\mathbf{k}_{\textrm{K}})=\pm| 3\sqrt{3}K_{0}(u)- 9K_{1}(u,\theta)|$
and
$\epsilon_{\pm}^{(\textrm{eff})}(\mathbf{k}_{\textrm{K}^{\prime}})=\pm|3\sqrt{3}K_{0}(u)+ 9K_{1}(u,\theta)|$,
where
$\mathbf{k}_{\textrm{K}}=(\frac{4\pi}{3\sqrt{3}}\ 0)^{T}$
and $\mathbf{k}_{\textrm{K}^{\prime}}=(\frac{8\pi}{3\sqrt{3}}\ 0)^{T}$;
those for $\beta=3$ are 
$\epsilon_{\pm}^{(\textrm{eff})}(\mathbf{k}_{\textrm{K}})=\epsilon_{\pm}^{(\textrm{eff})}(\mathbf{k}_{\textrm{K}^{\prime}})$.
Therefore, 
the valley degeneracy can be lifted for even $\beta$
due to a combination of breaking both time-reversal symmetry and inversion symmetry,
whereas it is preserved for odd $\beta$.

The energy difference between the two valleys,
$\Delta_{\textrm{valley}}
=|\epsilon_{\pm}^{(\textrm{eff})}(\mathbf{k}_{\textrm{K}})
-\epsilon_{\pm}^{(\textrm{eff})}(\mathbf{k}_{\textrm{K}^{\prime}})|$,
for $\beta=2$ can be controlled by
changing $u$ and $\theta$.
Figures \ref{fig2}(a){--}\ref{fig2}(c)
show
the $\theta$ dependence of $\epsilon_{\pm}^{(\textrm{eff})}(\mathbf{k})$ numerically calculated 
for $\beta=2$ at $\Omega=10t_{\textrm{NN}}$ and $u=0.4$ and $0.8$.
(For details of the numerical calculations, see the Supplemental Material~\cite{SM}.)
At $\theta=0$ and $\pi$,
$\Delta_{\textrm{valley}}$ is finite and increases with increasing $u$
[see Figs. \ref{fig2}(a) and \ref{fig2}(c)].
This is because
$K_{0}(u)$ and $|K_{1}(u,\theta)|$ increase with increasing $u$.
The valley which has the larger gap can be switched 
by changing $\theta$ from $0$ to $\pi$ or vice versa
[compare Figs. \ref{fig2}(a) and \ref{fig2}(c)]. 
Meanwhile,
at $\theta=\frac{\pi}{2}$,
$\Delta_{\textrm{valley}}=0$, i.e., the valley degeneracy holds [see Fig. \ref{fig2}(b)].
This is because $K_{1}(u,\theta)\propto \cos\theta$. 

\textit{Valley-selective Hall effect.}We turn to
the nonperturbative effects of BCPL on the Berry curvatures
for the driven system described by Eq. (\ref{eq:Heff}) for $\beta=2$.
The Berry curvature for the upper or lower band,
$B_{+}^{z}(\mathbf{k})$ or $B_{-}^{z}(\mathbf{k})$,
is given by
\begin{align}
  B_{\pm}^{z}(\mathbf{k})
  =-i\frac{v_{\pm\mp}^{x}(\mathbf{k})v_{\mp\pm}^{y}(\mathbf{k})
    -v_{\pm\mp}^{y}(\mathbf{k})v_{\mp\pm}^{x}(\mathbf{k})}
  {[\epsilon_{+}^{(\textrm{eff})}(\mathbf{k})-\epsilon_{-}^{(\textrm{eff})}(\mathbf{k})]^{2}},
\end{align}
where $v_{\alpha\beta}^{\nu}(\mathbf{k})
=\sum_{a,b=A,B}(U_{\mathbf{k}}^{\dagger})_{\alpha a}v_{ab}^{(\textrm{eff})\nu}(\mathbf{k})(U_{\mathbf{k}})_{b\beta}$
($\alpha,\beta=+,-$ and $\nu=x,y$),
$(U_{\mathbf{k}})_{a\alpha}$ is the unitary matrix to diagonalize Eq. (\ref{eq:Heff}),
and $v_{ab}^{(\textrm{eff})\nu}(\mathbf{k})
=\frac{\partial \bar{\epsilon}_{ab}(\mathbf{k})}{\partial k_{\nu}}$.
Figures \ref{fig2}(d){--}\ref{fig2}(f)
show the $\theta$ dependences of $B_{+}^{z}(\mathbf{k})$ and $B_{-}^{z}(\mathbf{k})$
numerically calculated
for $\beta=2$ at $\Omega=10t_{\textrm{NN}}$ and $u=0.4$ and $0.8$. 
(For details of the numerical calculations, see the Supplemental Material~\cite{SM}.)
If the valley degeneracy is lifted (i.e., $\theta$ is $0$ or $\pi$), 
the Berry curvatures at the two valleys are different in magnitude.
The valley which gives the largest contribution to the Berry curvatures
can be switched by changing $\theta$ from $0$ to $\pi$
or vice versa.
These results suggest that
BCPL could induce the valley-selective Hall effect
and switch its dominant valley by changing $\theta$.
Furthermore, by changing $\theta$
from $0$ or $\pi$ to $\frac{\pi}{2}$ or vice versa,
a crossover between
the valley-selective and valley-independent Hall effects could be induced.
However,
even with broken inversion symmetry,
the Berry curvatures of the upper or lower band at the two valleys
have the same sign, which means the absence of the valley-contrasting Hall effect. 

The results shown above are consistent with
the time-averaged anomalous Hall conductivity (AHC) $\sigma_{yx}^{\textrm{C}}$ calculated
in the Floquet linear-response theory~\cite{Eckstein,Oka-PRB,Tsuji,Mikami,NA-FloquetSHE,NA-linear}.
Figure \ref{fig3}(a) or \ref{fig3}(b)
shows $\sigma_{yx}^{\textrm{C}}$ numerically calculated at $\Omega=8t_{\textrm{NN}}$
for $\beta=2$ or $3$.
(For details of the numerical calculations, see the Supplemental Material~\cite{SM}.)
For $\beta=3$,
$\sigma_{yx}^{\textrm{C}}$ is quantized in a similar way to
that of graphene driven by CPL~\cite{Oka-PRB,Mikami,Kitagawa}.
This agrees with the analyses using the high-frequency expansion
because the effective Hamiltonian for $\beta=3$ is qualitatively the same as
that of graphene driven by CPL~\cite{Mikami,Kitagawa}.
Meanwhile,
for $\beta=2$,
$\sigma_{yx}^{\textrm{C}}$ is quantized
at $\theta=\frac{\pi}{2}$,
whereas that is reduced from the quantized value for moderately large $u$'s
at $\theta=0$ and $\pi$.
The magnitude reduction in $\sigma_{yx}^{\textrm{C}}$
for large $u$'s at $\theta=0$ or $\pi$
is attributed to a drastic reduction in the contributions near one valley
due to a larger gap opening.
In all the cases,
$\sigma_{yx}^{\textrm{C}}\approx -\sigma_{xy}^{\textrm{C}}$ is satisfied
(see Fig. 2 in the Supplemental Material~\cite{SM}).
(Because of this,
it is reasonable to call $\sigma_{yx}^{\textrm{C}}$ the anomalous Hall conductivity.)
Since 
such an antisymmetric part
$(\sigma_{yx}^{\textrm{C}}-\sigma_{xy}^{\textrm{C}})/2$ is finite
only with broken time-reversal symmetry~\cite{Onsager1,Onsager2},
these results indicate that
time-reversal symmetry is broken in all the cases, which agrees with
the high-frequency expansion.

Then, 
Figs. \ref{fig3}(c){--}\ref{fig3}(h) show 
$\sigma_{yx}^{\textrm{C}}(\mathbf{k})$'s
calculated in the Floquet linear-response theory
for $\beta=2$ at $\Omega=8t_{\textrm{NN}}$
with $(\theta,u)=(0,0.2)$, $(0,0.4)$, $(0,0.6)$,
$(\pi,0.2)$, $(\pi,0.4)$, and $(\pi,0.6)$,
where $\sigma_{yx}^{\textrm{C}}(\mathbf{k})$ is defined as
$\sigma_{yx}^{\textrm{C}}=\sum_{\mathbf{k}}\sigma_{yx}^{\textrm{C}}(\mathbf{k})$.
For $u=0.2$,
$\sigma_{yx}^{\textrm{C}}(\mathbf{k}_{\textrm{K}})$
and $\sigma_{yx}^{\textrm{C}}(\mathbf{k}_{\textrm{K}^{\prime}})$ are almost the same.
Meanwhile,
for $u=0.4$ and $0.6$,
the main contribution at $\theta=0$ or $\pi$
comes from the vicinity at the K or K$^{\prime}$ point, respectively. 
These results indicate that
moderately strong BCPL can induce 
the valley-selective Hall effect,
and that
the valley which gives the main contribution to this Hall effect
can be switched by changing $\theta$
from $0$ to $\pi$ or vice versa.
In contrast,
$\sigma_{yx}^{\textrm{C}}(\mathbf{k}_{\textrm{K}})$
and $\sigma_{yx}^{\textrm{C}}(\mathbf{k}_{\textrm{K}^{\prime}})$
are degenerate at $\theta=\frac{\pi}{2}$ even for $u=0.4$ and $0.6$
(see Fig. 3 in the Supplemental Material~\cite{SM});
the same degeneracy holds for $\beta=3$
at $\theta=0$, $\frac{\pi}{2}$, and $\pi$ (see Fig. 3 in the Supplemental Material~\cite{SM}).
Therefore,
by changing $\theta$ for $\beta=2$,
the crossover between the valley-selective and valley-independent Hall effects
can be induced.
Since the valley degeneracy is lifted only
without both time-reversal symmetry and inversion symmetry,
these results
indicate that
inversion symmetry is broken for even $\beta$ at $\theta=0$ and $\pi$,
which also agrees with the high-frequency expansion.

Similar results are obtained at $\Omega=6t_{\textrm{NN}}$ 
(see Fig. 4 in the Supplemental Material~\cite{SM}),
implying that
the similar properties hold even in the resonant case.
Therefore,
our valley-selective Hall effect could be experimentally observed
because for graphene driven by CPL,
the results obtained in the Floquet linear-response theory~\cite{Oka-PRB,Mikami}
are qualitatively reproducible in experiments 
using a smaller light frequency~\cite{Light-AHE-exp2}.
Note that
the Floquet linear-response theory can analyze
the off-resonant case and the resonant case,
whereas the high-frequency expansion can analyze only the former case.
Our light is resonant or off-resonant
if $\Omega \leq W_{\textrm{band}}$ or $\Omega > W_{\textrm{band}}$,
respectively,
where $W_{\textrm{band}}=6t_{\textrm{NN}}$ is the bandwidth of the nondriven system.  

\begin{figure}
  \includegraphics[width=86mm]{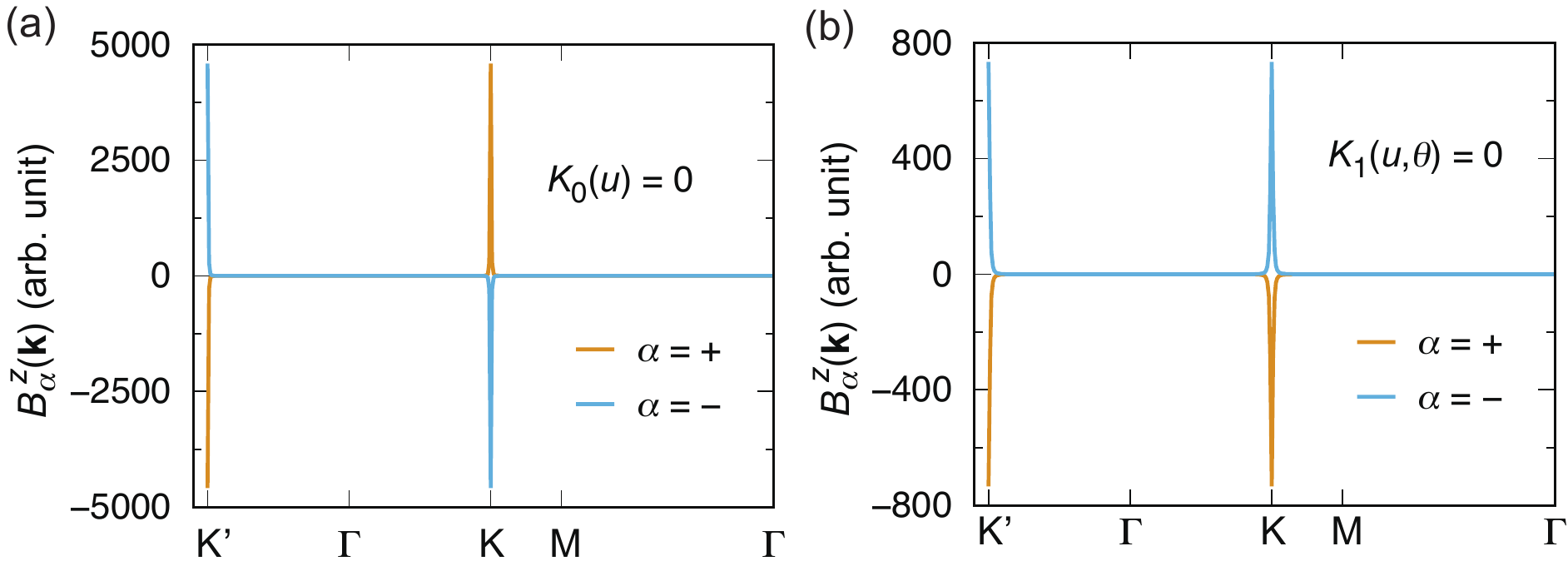}
  \caption{\label{fig4}
    (a), (b) The momentum dependences of $B_{+}^{z}(\mathbf{k})$ and $B_{-}^{z}(\mathbf{k})$ 
    in the special cases with $K_{0}(u)=0$ and $K_{1}(u,\theta)=0$.
    The symmetric points are the same as those used in Fig. \ref{fig2}.
    }
\end{figure}

\textit{Discussion.}To understand
why the valley-contrasting Hall effect is absent, 
we compare the Berry curvatures calculated for $\beta=2$ using the high-frequency expansion
with those in two special cases.
In these cases, 
the effective Hamiltonian is given by Eq. (\ref{eq:Heff}) for $\beta=2$ 
with $K_{0}(u)=0$ or $K_{1}(u,\theta)=0$, 
and the other parameters are the same as those used
for $\beta=2$ and $\Omega=10t_{\textrm{NN}}$ at $\theta=0$ and $u=0.4$.
Since the $K_{0}(u)$ term and $K_{1}(u,\theta)$ terms break
time-reversal symmetry and inversion symmetry,
respectively [Eqs. (\ref{eq:Del-beta}) and (\ref{eq:K-beta})], 
the first case possesses time-reversal symmetry,
whereas the other possesses inversion symmetry.
(The first case is similar to that studied
in Ref. \onlinecite{VHE-theory}.)
Note that for $\beta=2$,
$K_{1}(u,\theta)$ appears in 
$\Delta_{\textrm{BCPL}}$ and $K_{\textrm{eff}}^{(AA)}(\pm\mathbf{R}_{l}^{\prime})$
[Eqs. (\ref{eq:Del-beta}) and (\ref{eq:K-beta})].
Figures \ref{fig4}(a) and \ref{fig4}(b)
show the Berry curvatures numerically calculated in these two cases.
The Berry curvatures of the upper or lower band at the two valleys
are opposite in sign with time-reversal symmetry.
Therefore, 
we conclude that
the valley-contrasting Hall effect is absent in graphene driven by BCPL
due to the broken time-reversal symmetry.
Note that
the difference between the signs of the Berry curvatures at the two valleys
without and with time-reversal symmetry
may be similar to that between a ferromagnet and an antiferromagnet. 

We compare our study with the relevant studies. 
The BCPL-induced symmetry breaking has been partly clarified
in some off-resonant cases at $\beta=2$
using the high-frequency expansion~\cite{Bicirc-Oka,Bicirc-PRL}.
Meanwhile, our results obtained in the high-frequency expansion are applicable
to any $\beta$.
Furthermore, our Floquet linear-response theory showed 
that these results remain qualitatively unchanged
at smaller light frequencies including a resonant one for $\beta=2$ and $3$. 
Therefore, our results have a wider applicability. 
Then,
there is no previous study showing the valley-selective Hall effect
via the light-induced inversion symmetry breaking,
although there are many studies about
valley-dependent properties of
periodically driven systems~\cite{Circ2-Floquet-graphene,BCPL-Floquet-graphene1,BCPL-Floquet-graphene2,Valley-Abergel,staticE-ISB1,staticE-ISB2}.
Therefore, our study is the first work demonstrating
the BCPL-induced valley-selective Hall effect.

We also comment on three advantages of our valley-selective Hall effect.
In the standard mechanism~\cite{VHE-exp1,VHE-exp2},
the inversion symmetry is broken by the lattice
and the time-reversal symmetry is broken by resonant CPL.
Meanwhile, in our mechanism, 
these symmetries are both broken by BCPL. 
Therefore, 
only our mechanism 
works in centrosymmetric systems.
In addition,
our mechanism enables us to switch the main valley
and induce the crossover between the valley-selective and valley-independent Hall effects
by tuning $\theta$.
Note that in the standard mechanism,
the main valley can be switched by changing the helicity of CPL. 
Then,
our Hall conductivity, which
is $O(e^{2}/h)$ [e.g., see the value at $u=0.6$ and $\theta=0$ in Fig. \ref{fig3}(a)],
is two orders of magnitude larger than
the maximum value obtained in the standard mechanism~\cite{VHE-exp2}.
This comes from a special property of monolayer graphene that 
the energy gaps at the two valleys are tunable solely by light,
which enables us to make 
the energy gap at one valley much larger than the other
with keeping the other small.
Therefore, monolayer graphene driven by BCPL may provide
the best opportunity for the valley-selective Hall effect.

\section{Acknowledgments}

This work was supported by
JST CREST Grant No. JPMJCR1901, 
JSPS KAKENHI Grant No. JP22K03532, 
and MEXT Q-LEAP Grant No. JP-MXS0118067426.

\end{document}